# Mechanics of cell integration *in vivo*

Guilherme B. Ventura[1,*], Aboutaleb Amiri[2,*], Raghavan Thiagarajan[1], Mari Tolonen[1], Amin Doostmohammadi[3,†], Jakub Sedzinski[1,†]

**Affiliations:**

[1]The Novo Nordisk Foundation Center for Stem Cell Biology, University of Copenhagen, Blegdamsvej 3B, 2200 Copenhagen, Denmark

[2]Max Planck Institute for the Physics of Complex Systems, Nöthnitzer Str. 38, 01187 Dresden, Germany

[3]The Niels Bohr Institute, University of Copenhagen, Blegdamsvej 17, 2100 Copenhagen, Denmark

* co-first authors

†Corresponding author. Email: jakub.sedzinski@sund.ku.dk (J.S.); doostmohammadi@nbi.ku.dk (A.D.).

**ABSTRACT**

*During embryonic development, regeneration and homeostasis, cells have to physically integrate into their target tissues, where they ultimately execute their function. Despite a significant body of research on how mechanical forces instruct cellular behaviors within the plane of an epithelium, very little is known about the mechanical interplay at the interface between migrating cells and their surrounding tissue, which has its own dynamics, architecture and identity. Here, using quantitative in vivo imaging and molecular perturbations, together with a theoretical model, we reveal that multiciliated cell (MCC) precursors in the Xenopus embryo form dynamic filopodia that pull at the vertices of the overlying epithelial sheet to probe their stiffness and identify the preferred positions for their integration into the tissue. Moreover, we report a novel function for a structural component of vertices, the lipolysis-stimulated lipoprotein receptor (LSR), in filopodia dynamics and show its critical role in cell intercalation. Remarkably, we find that pulling forces*

*equip the MCCs to remodel the epithelial junctions of the neighboring tissue, enabling them to generate a permissive environment for their integration. Our findings reveal the intricate physical crosstalk at the cell-tissue interface and uncover previously unknown functions for mechanical forces in orchestrating cell integration.*

**MAIN**

As cells migrate through or push and pull on their neighbors in tissues, they are embedded in a complex 3D environment that continuously exposes them to diverse mechanical stimuli[1]. The combination of biophysical and theoretical methods together with recent advances in measuring mechanical stresses *in vivo* has revealed how cells mechanically interact with their passive environment, for example, by sensing the stiffness of the extracellular matrix (ECM)[2–4], and how mechanical forces drive cellular behaviors in the plane of the epithelial monolayers[5]. Despite these advances, we know comparatively little about the mechanical crosstalk at the interface of migrating cells and the surrounding tissues, which underlies a range of developmental, regenerative and pathological events, e.g., during epithelialization, homeostatic cell renewal and cancer cell invasion[6–9].

Common to many of these cell-tissue interactions is the movement and subsequent integration of cells within the overlying epithelium. During the formation of the mucociliary epithelium in the amphibian *Xenopus* embryo, successive waves of precursor cells move from the basal into the superficial tissue layer[10–13]. The first wave of migrating cells is composed of multiciliated cell (MCC) precursors, which integrate into the superficial epithelial layer composed of mucus-producing goblet cells[10,14,15] (Fig. 1a). This multistep process, collectively known as radial intercalation, requires a complex cell-tissue interplay (Supplementary Fig. 1a). Previous studies have shown that MCCs move into the overlying epithelium and insert at the epithelial vertices formed by three cells, commonly referred to as tricellular junctions[16–19] (Fig. 1a,b and Supplementary Fig. 1a). Thus, these structures act as the susceptible positions within the epithelium that facilitate the most efficient integration within the tissue[10].

Epithelial vertices have recently been identified as key structural components integrating and controlling both biochemical and mechanical cues within epithelial sheets and as being responsible for directing cell division and cell migration[20–22]. Of particular interest, epithelial vertices have been described as mechanical hotspots within the tissue as they sustain the tension generated by the connecting epithelial junctions[23–25] (Fig. 1a, inset). However, how this distinct mechanical feature of epithelial vertices contributes to cell integration and how radially intercalating cells select which vertex to insert into is unknown.

**Multiciliated cells form dynamic filopodia targeting the epithelial vertices**

To explore the potential mechanical crosstalk between the epithelial vertices and the intercalating cells, we first characterized the dynamics of migrating MCCs as they begin to move into the superficial epithelial layer[10]. Using cell-type-specific α-*tubulin* and *nectin* promoters[14], we expressed actin biosensors LifeAct-GFP and Utrophin-RFP in the MCCs and the neighboring goblet cells, respectively. Three-dimensional (3D) time-lapse imaging revealed that MCCs accumulated filamentous actin (F-actin) at their leading-edge, from which they extended finger-like protrusions as they ascended apically. These F-actin-rich filopodia were dynamic and pointed at the cell junctions overlaying the MCCs (Figs. 1b,c; Supplementary Fig. 1b and Supplementary Video 1). We observed that during this behavior, MCCs interacted with multiple vertices as they moved laterally unrestricted from other neighboring MCCs (Figs. 1d-e, Supplementary Fig. 1b and Supplementary Video 1). To understand whether filopodia are randomly assembled along the leading edge or if they are directed to specific positions, such as the epithelial vertices, we established an image analysis pipeline to quantify F-actin protrusion activity and position within the MCC's leading edge during intercalation (Fig. 1f and Supplementary Fig.1c-d). Our analysis revealed that while cells extended filopodia along their entire leading edge, filopodia were consistently enriched at vertices (Fig. 1g and Supplementary Video 2). Remarkably, we observed that cells did not interact with one single vertex at the time, but extended filopodia at several vertices in the cell's vicinity and often moved closer to a neighboring vertex after its initial probing (Fig. 1g, Supplementary Fig.1e and Supplementary Video 1). Combined, these results show that filopodia are consistently formed at the positions of the leading edge closest to the vertices and that they are coordinated with MCC axial and lateral movement, strongly suggesting that the filopodia guide cell movement by transmitting spatial information from the surrounding goblet cells. Interestingly, filopodia are known to exert pulling forces in the ECM to probe its mechanical properties[26,27]. Therefore, we asked whether MCC-generated filopodia have a similar role by sensing the mechanical features of the overlying epithelial vertices.

**LSR mediates direct interactions between filopodia and epithelial vertices**

To further examine the relationship between the filopodia generated by the MCCs and the epithelial vertices, we imaged one of the main structural components of the epithelial vertices, the lipolysis stimulated lipoprotein receptor (LSR/angulin-1)[28]. LSR extends basolaterally to form a string-like structure[29] (Supplementary Fig. 2a). We visualized LSR-3xGFP[30] in the superficial cells and performed dynamic imaging of intercalating MCCs expressing LifeAct-RFP (Fig. 2a-c and Supplementary Video 3). We confirmed that the vertices were consistently targeted by dynamic filopodia which formed temporary contacts with the LSR strings (Fig. 2a-c, S2a and Supplementary Video 3). Upon multiple cycles of contact formation and retraction between filopodia with the LSR string, filopodia concentrated around the LSR structure, leading to the accumulation of F-actin within the MCC cortex at the selected vertex (Fig. S2d,f). These results reinforce the notion that a close contact between the MCC and the vertices, mediated by the filopodia, precedes cell insertion.

We surprisingly found that LSR also localized at the tips of filopodia, where it was maintained during filopodia extension and retraction (Fig. 2d, Supplementary Fig. 2d,g and Supplementary Video 4). We then hypothesised that MCCs could recruit LSR to the leading edge to form LSR-LSR contacts with the vertices. Using cell-type specific expression of LSR, we observed that such homotypic LSR-LSR connections exist (Supplementary Fig. 2e). Thus, our data demonstrate that MCCs use filopodia to directly interact with vertices by LSR-LSR mediated contacts.

We observed that LSR knockdown in MCCs dramatically impaired radial intercalation by blocking apical emergence (86% of all LSR MO cells) (Supplementary Fig. 3a,b and Supplementary Video 5). Notably, LSR depleted MCCs were able to reach the superficial epithelium, but some disappeared back into the basal layer, suggesting that a strong attachment to the overlying vertices might be required to stabilize the MCCs in the superficial layer (27% of all LSR MO cells) (Supplementary Fig. 3c). LSR depleted cells also failed to sustain any prominent filopodia growth, in contrast with the control cells (Supplementary Fig. 4a-f). Remarkably, we also observed that LSR overexpression in MCCs induced the ectopic formation of filopodia from expanding apical domains (Supplementary Fig. 2h and Supplementary Video 6). Combined with the observation that LSR mediates the interaction between filopodia and vertices, these findings show that LSR regulates the actin cortex dynamics and filopodia activity in the MCCs, which are in turn required for successful cell integration within the epithelium (Supplementary Fig. 2i).

In addition to dissecting the mechanism of interaction between the intercalating MCC and the epithelial vertex, we observed that as filopodia form contacts with the LSR string, they are able to pull on the epithelial vertices (Fig. 2f, Supplementary Fig. 2a,b and Supplementary Video 7). Upon pulling by filopodia, the vertex underwent a quick elongation followed by retraction as the filopodia detached (Fig. 2e,f, Supplementary Fig. 2a-c and Supplementary Video 8). We hypothesized that MCCs could exploit the ability to pull on the vertices of the overlying epithelium to probe for points in the tissue which can be exploited for their intercalation (Fig. 2g).

**MCCs probe local vertex stiffness in the overlying epithelium**

To provide a quantitative understanding of how MCCs could probe the mechanical landscape of their overlying epithelium, we developed a minimal theoretical model of cell-tissue interaction based on a vertex-based model[31] (see Supplementary Section 1). To simulate filopodia-induced pulling in a tissue of heterogeneous line tension, we sequentially applied an out-of-plane force of a fixed magnitude, $f$, at each vertex, while maintaining all other vertices in the plane and measured the out-of-plane displacement, $\delta$, of the vertex as a result of the applied force (Fig. 3a). By repeating this step for all the vertices in the epithelial sheet, we obtained the map of local *vertex stiffness* $K_\delta = f/\delta$ (Fig. 3a,b) and found that out-of-plane pulling can effectively distinguish vertices with connecting junctions of high line tension from those with low tension connectivity (Fig. 3b, see Supplementary Model Section 1). We further explored the susceptibility of different vertices to intercalation by inserting a cell of an initial area $A_{in}$, much smaller than the preferred cell area $A_0$, at each of the vertices, one at a time. In this model, the pressure difference between the inserting cell and its neighbors expands the cell toward the target area, which is resisted by the inserting cell's cortical tension and enhanced by line tensions of connecting junctions constituting the vertex[14]. The model predicted a high propensity of successful intercalations at the vertices where local stiffness is greatest (Fig. 3c-e). This is due to the overall higher line tension of the adjacent cell junctions pulling at the vertex, which enhances the expansion of the MCC's apical domain within the epithelial sheet (Fig. 3b,c). Moreover, the existence of heterogeneous line tension can induce the formation of higher fold vertices, where 4- or more cells meet[32] and the model predicted that 4- and higher-fold (high-fold) vertices should open up easier than the predominant 3-fold vertices in the overlying tissue (Fig. 3f) (see Supplementary Section 1).

To test the model predictions experimentally, we quantified the evolution of vertex fold number, vertex stiffness and the propensity of MCCs to intercalate at a particular vertex type. Informed by the *in silico* model prediction that suggests vertex stiffness directly depends on the tension of junctions connecting the vertex (Fig. 3b), we measured the sum of line tensions of the junctions connected to vertices as a proxy of vertex stiffness *in vivo*. Specifically, as a readout of junctional tension, we quantified the myosin intensity using a myosin II-specific nanobody (SF9-3xGFP)[33] (Fig. 3g,h and Supplementary Video 9). Consistent with our model's predictions, we found that vertex stiffness scaled up with the vertex fold number and that it remained relatively constant throughout radial intercalation (Fig. 3i). Additionally, we manually scored the timing of MCC insertion and quantified the number of neighbours to see if MCCs prefer higher-fold intercalations (Fig. 3j-k and Supplementary Video 9). We found that, cumulatively, intercalations at 5-, 6- and 7-fold vertices accounted for 36.2% of all intercalation events; 4-fold intercalations accounted for 57.1% of all events whereas 3-fold intercalations represented only 6.6% of all intercalations, consistent with previous work[19] (Fig. 3l-m). Thus our experimental results validate the prediction from the model that MCCs preferentially intercalate at high-fold vertices that accommodate higher tension connectivity (Fig. 3m). Nevertheless, it remains unclear how these entry points are formed. The quantitative analysis of the movies revealed a decrease in the number of 3-fold vertices and the concomitant formation of 4-fold vertices, which interestingly preceded the onset of MCCs insertion (Fig. 3n). This suggests that there is an active mechanism that promotes the formation of the 4-fold vertices which are preferred insertion points.

When we quantified the distance between the position of a junction collapse and an MCC intercalation event, we observed that MCCs predominantly intercalated in the immediate vicinity of the collapsed junction (Fig. 3o and Supplementary Fig. 5a,b). Analysis of the time-lapse movies supported this observation as 87% (n=147 of 168 events) of MCC intercalation events in the high-fold vertices coincided with cell junctions remodelling leading to the conversion of three-fold vertices into high-fold vertices (Fig. 3p and Supplementary Fig. 5c,d). Moreover, we found that the start of junction remodelling was only observed when MCCs reached the superficial layer (Supplementary Fig. 5e).

**MCCs actively remodel epithelia to induce higher-fold vertices**

To address the possibility that junction remodelling is dependent on the integrating MCC, we explored the mechanisms underlying the emergence of high-fold vertices. Given that junction remodelling has been extensively described to be driven by cell junction contraction and reliant on the activity of non-muscle myosin II[34–36], we first tested whether the formation of higher-fold vertices is driven by myosin II in the goblet cells (Supplementary Fig. 6a). Surprisingly, we observed no evident accumulation of myosin prior to junction remodelling (Fig. 4a,b; Supplementary Fig. 6b and Supplementary Video 10). Instead, we found that myosin was only accumulated after the junction starts shrinking, suggesting that myosin II accumulated in response to an external stimulus that promotes the initiation of junction remodelling (Fig. 4b, Supplementary Fig. 5b and Supplementary Video 10). To study this intriguing possibility, we next analyzed the relative position of the intercalating cell at the onset of junction shrinking. Interestingly, junction shrinking followed the formation of stable contacts between the MCC and the vertices (Fig. 4a,c,e and Supplementary Fig. 6i,j). Moreover, using F-actin intensity in the MCC as a proxy for the proximity of the MCC to the overlying junction, we observed that the increase in the MCC F-actin intensity preceded both the start of junction shortening and the accumulation of junctional myosin II in the goblet cells (Fig. 4c,g, and Supplementary Fig. 6c,d). This, together with our data on the onset of junction remodelling, suggests that MCCs initiate the active remodelling of the superficial epithelial layer, however, it does not fully explain how the intercalating cells achieve this.

To address this question, we further envisioned two alternative mechanisms: i) indirect, where MCCs induce the goblet cells to shrink the junctions after establishing close contact, and ii) direct, where continuous pulling at the vertices by the MCC enforces the neighbor remodelling. To distinguish between these two alternative mechanisms, we reasoned that if the vertices are actively being pulled by the intercalating MCC, then we should observe quick vertex retraction whenever the interacting cell loses contact. To test this hypothesis, we tracked junction remodelling in the early stages of intercalation, in which cells are able to freely interact with multiple vertices. We observed that junction shrinking could be quickly reverted whenever the intercalating MCC lost direct contact with one of the overlying vertices (Supplementary Fig. 6e-k, Supplementary Videos 11 and 12).

To further understand how this novel pulling mechanism by the intercalating cells could trigger junction remodelling, we returned to our theoretical model and conducted a numerical experiment in which we probed the response of junctions - one at a time - to external contraction (Fig. 4h). In order to quantify the junction fate under tensional perturbation, we define $l_f / l_i$ as the order parameter characterizing the ratio of the particular junction's length after applying a positive perturbation to its initial junctional tension (Fig. 4h). The simulation results showed that 4-fold vertices were formed from the collapse of junctions with sufficiently short length and sufficiently large tension (Fig. 4h). A qualitatively similar phase diagram is observed for the experimental data (Fig. 4i). Moreover, compared to the line tension, the initial length of the junction plays a more dominant role in determining whether the junction collapses or not, as the order parameter is more sensitive to changes in the initial junction length (Fig. 4h-j). Combined, these results show that when intercalating MCCs enhance the effective tension of a junction in the superficial layer by pulling on its two vertices, they can indeed induce a re-arrangement in the overlying layer by shrinking that junction and forming a 4-fold vertex, which in turn is a preferential site for the MCCs to insert themselves. Therefore, MCCs can create their own 4-fold vertex by inducing junction remodelling, if the junction is sufficiently short and its tension is sufficiently large.

Altogether, our results suggest that, after probing the neighboring tissue environment, MCCs induce junction shrinkage in a multi-step process (Fig. 4k). First, cells establish stable attachments with surrounding vertices. The intercalating MCCs are then capable of pulling on the overlying junction and initiate its remodelling. Subsequently, myosin II accumulation at the junctions by the goblet cells reinforces junction collapse. Thus, intercalating MCCs exert forces on an active pliable environment to create advantageous insertion points.

**Myosin activity in the MCCs is required for junction remodelling and intercalation**
We next explored how the intercalating MCCs exert pulling forces to remodel the neighboring environment and produce favorable insertion points. A prime candidate for this would be myosin II, a key force-generating molecular motor[27], so we decided to express SF9-3xGFP mRNA, this time in MCCs. We observed that myosin localized to the leading edge of intercalating cells, where it first appeared at the base of the filopodia (Fig. 5a-b, Supplementary Fig. 7a-b and Supplementary Video 13). We then quantified how myosin is recruited to the leading edge of the intercalating cells which have already formed contacts with the vertices and observed that myosin was progressively enriched similarly to F-actin (Supplementary Fig. 7c-d). To test whether the MCCs' ability to intercalate depends on their capacity to exert forces on their neighbors, we assessed the impact of myosin II inhibition on MCCs intercalation. To this end, we mosaically expressed a constitutively active form of the myosin light chain phosphatase (CA-MLCP) in the MCCs, which dephosphorylates myosin II and negatively regulates actomyosin contractility[4] (Fig. 5c). Our dynamic imaging experiments verified that CA-MLCP expressing MCCs failed to both intercalate and to remodel the overlying junctions (Fig. 5c-f, Supplementary Fig. 7e and Supplementary Video 14). Together, the theoretical and experimental data demonstrate that MCCs autonomously initiate and then cooperatively drive the remodelling of epithelial junctions to facilitate epithelial integration.

**DISCUSSION**

Here, using radially intercalating MCCs *in vivo* as a model system, we have begun to dissect the mechanisms regulating the mechanical interplay at the interface of an integrating cell and the surrounding tissue. Our experimental and theoretical results identify actin-based filopodia and vertices as the main players involved in probing tissue mechanics and put forward a novel concept of how a single cell can sense, interpret and remodel the neighboring cellular microenvironment to coordinate its behavior.

Our findings show, for the first time, that during radial intercalation, hundreds of migrating MCCs actively decide in which vertex to insert. This choice does not follow a straightforward "first-come-first-serve" principle. Cell intercalation is instead guided, as the migrating MCCs use filopodia to actively and directly sense the vertices of the overlying epithelium. Impairing filopodia activity leads to radial intercalation defects as MCCs fail to anchor to the vertices. Therefore, we propose that filopodia provide the intercalating MCCs with a parallel guiding mechanism to the recently described Scf/kit biochemical signalling pathway[37]. While Scf/kit controls both the spacing between neighboring MCCs and their overall affinity to the overlying epithelium, filopodia precisely inform the MCCs on the position of the vertices.

Our work further adds LSR as an important regulator and potential mechanotransducer involved in cellular microenvironment sensing. First, we reveal a previously uncharacterized localization of LSR to the tips of filopodia. Secondly, we find that LSR depletion in MCCs blocks filopodia formation, impairs cortical actin assembly and leads to radial intercalation defects. Finally, we show that myosin II is recruited to the base of filopodium and its downregulation leads to the impairment of MCCs radial intercalation, similarly to LSR knockdown. Altogether, our results describe the mechanistic basis for how filopodia are required for proper MCC integration within the overlying tissue.

Importantly, our work addresses the fundamental question of how MCCs select a particular insertion point. A recent study has described the propensity of MCCs toward higher-fold vertices and that increasing a cell's capability to insert by promoting microtubule acetylation skews the intercalation propensity toward lower-fold vertices[19]. Nonetheless, why MCCs preferentially insert at higher-fold vertices remains an outstanding question. Our work addresses this unresolved phenomenon, and we propose that the cell's decision in which vertex to intercalate is ultimately determined by its stiffness. Interestingly, probing by filopodia is known to be used by migrating cells to measure temporal variations of local ECM stiffness *in vitro*[38]. We show that an out-of-plane pulling force can effectively probe vertex stiffness to identify an ideal vertex: one in which high enough line tension promotes vertex opening. Together, our results provide the first insights into the basis of mechanical probing of tissues by filopodia *in vivo*.

Strikingly, our findings reveal that MCCs are able to remodel the overlying cell-cell junctions to form the higher-order vertices in which they predominantly intercalate. Junction remodelling is known to be the driver of many morphological processes in epithelial tissues[39,40]. Until now, these processes have been largely characterized to be driven by cellular forces generated within the epithelial layer[5]. However, our results show that intercalating MCCs are also able to promote similar remodelling to create the optimal mechanical environment for their integration within the tissue. This novel process relies on initial mechanical stimuli by the intercalating MCCs, which then cooperate with goblet cells to ensure complete junction remodelling.

Beyond the direct implications of our findings in understanding the fundamental concepts of cell integration *in vivo*, the experimental and mathematical framework described here provide novel insights into how cells sense and transmit mechanical cues from their environment to the cytoskeletal machinery that ultimately drives cellular behavior, a deeply fundamental question in biology. Finally, as both filopodia and mechanical stimuli are involved in a plethora of processes, from development to cancer [41–43], we expect our work to help understand how these two essential players are intertwined to guide cell behavior in both normal and pathological conditions.

**Acknowledgments:**

The authors thank all members of the Sedzinski lab and Active & Intelligent Matter Group members for comments and suggestions, Elke Ober and Mariaceleste Aragona for critical reading of the manuscript, Ann L. Miller, Edwin Munro and Elias H. Barriga for key reagents and Benoit Aigouy for help with Tissue Analyzer. We acknowledge the Core Facility for Integrated Microscopy, Faculty of Health and Medical Sciences, University of Copenhagen and DanStem's microscopy specialist, Jutta M. Bulkescher for training, technical expertise, support, and microscope use.

**Funding**

The Novo Nordisk Foundation Center for Stem Cell Biology is supported by a Novo Nordisk Foundation grant number NNF17CC0027852; J.S. acknowledges support from the Novo Nordisk Foundation (grant No. NNF19OC0056962); A.D. acknowledges support from the Novo Nordisk Foundation (grant No. NNF18SA0035142), Villum Fonden (Grant no. 29476), and funding from the European Union's Horizon 2020 research and innovation program under the Marie Sklodowska-Curie grant agreement No. 847523 (INTERACTIONS). A.A. acknowledges support from the Federal Ministry of Education and Research (Bundesministerium für Bildung und Forschung, BMBF) under project 031L0160.

**Author contributions**

G.B.V. performed all the experiments, analyzed the data and prepared the figures; M.K.T helped with image segmentation; R.T. helped with image analysis and developed the filopodia analysis pipeline; A.A. and A.D. developed the theoretical model and contributed to data analysis and figures; J.S., A.D., G.B.V., and A.A wrote the manuscript; All authors edited the manuscript; J.S. and G.B.V conceived the project. J.S. acquired funding and supervised the project.

**Competing interests**

The authors declare no competing interests.

## Data and materials availability

The data, code used for analysis, models and materials that support these findings are readily available from the corresponding authors, Jakub Sedzinski (jakub.sedzinski@sund.ku.dk) and Amin Doostmohammadi (doostmohammadi@nbi.ku.dk) upon request.

## Supplementary Materials:

## Materials and Methods

Key Resources Table

| Reagent or Resource | Source | Identifier |
|---|---|---|
| **Chemicals** | | |
| Ultra-Low Melting Point Agarose | Sigma Aldrich | A2576-5G |
| Abberior Start Phalloidin-488 | Abberior | |
| CasBlock™ | ThermoFisher | 008120 |
| | | |
| **Antibodies** | | |
| Anti-LSR | Higashi et al 2016 | |
| Donkey Anti-rabbit Alexa647 | ThermoFisher | A-31573 |
| | | |
| **Oligonucleotides** | | |
| Primers for cloning angulin-1 in α-tubulin backbone:<br><br>Fwd: | | |
| caccATGGAAGGGACCGGAATTG<br>Rev: AACGACCAAACTCTCTCGGC | | |
| | | |
| Primers for cloning angulin-1 in nectin backbone:<br><br>Fwd: aacggtGGATCCATGGAAGGGACCGGAATTG<br>Rev: aacggtGCGGCCGCAAAAAACCTCCCACACCTCCC | | |
| | | |
| Primers for cloning SF9 in pCS2+ 3xGFP backbone:<br><br>Fwd: aacggtATCGATATGGCCGAGGTGCAGCTGG<br>Rev: aacggtTCTAGAACCTAGGACGGTCAGCTTGGTC | | |
| | | |
| | | |
| **Recombinant DNA** | | |
| | | |
| pα-tubulin LifeAct-GFP | Sedzinski et al., 2016 | |
| pα-tubulin LifeAct-RFP | Sedzinski et al., 2016 | |
| pα-tubulin angulin-1-GFP | This study | |
| pNectin Utrophin-RFP | Sedzinski et al., 2016 | |
| pNectin angulin-1-RFP | This study | |

| pCS2+/LifeAct-GFP | Wallingford Lab | |
| pCS2+/H2B-RFP | Wallingford Lab | |
| pCS2+/angulin-1-3xGFP | Higashi et al., 2016 | |
| pCS2+/SF9-3xGFP | This study | |
| pCS2+/CA-MYPT | Barriga et al., 2019 | |
| pCS2+/CA-MLC-mCherry | Barriga et al., 2019 | |
| | | |
| **Oligomorpholinos** | | |
| LSR MO1: gagaacatggaagggaccggaattg | Gene Tools USA | |

## Xenopus Laevis Embryo Manipulation

*Xenopus Laevis* adult females were injected with human chorionic gonadotropin (Chorulon) to induce ovulation. Male frogs were sacrificed and their testis dissected for the sperm samples. *Xenopus Laevis* eggs were harvested, fertilized *in vitro* and dejellied with 3% cysteine (pH 7.9) solution after 2h. Cleaving embryos were then washed and reared in 1/3X Marc's Modified Ringer's (MMR) solution. For mRNA and plasmid microinjection, embryos were transferred to a 2% Ficoll in 1/3X MMR solution and injected using glass needles using a universal micromanipulator. The Danish National Animal Ethics Committee reviewed and approved all animal procedures (Permit number 2017-15-0201-01237).

## Construct preparation

The primers used for cloning are listed in the key resources table. Cloning of angulin-1 into the pα-tubulin backbone was performed using a combination of the pENTRE™-dTOPO (Thermo Fisher Scientific) and Gateway™ systems. The LSR coding sequence was PCR-amplified and inserted into a pENTRE™ vector by enzymatic reaction. The LSR CDS was then subcloned into a previously designed pα-tubulin by recombination using the Gateway™ LR clonase II enzyme mix (Thermo Fisher Scientific)[44]. Cloning of the SF9 myosin sensor into the pCS2+/3xGFP backbone was performed by restriction-ligation. All plasmids were confirmed by restriction and sequencing (Eurofins Genomics, DE).

**Construct synthesis and microinjection**

pCS2+ plasmids were linearized with NotI-HF, purified by gel-extraction and then used as templates for *in vitro* transcription. *In vitro* transcription of mRNAs was performed using the mMessage mMachine SP6 kit (Ambion). Synthesized mRNA was purified by LiCl precipitation. Plasmid and mRNA probes microinjected together or separately into 4-cell stage embryos with a single injection into each ventral blastomere. mRNA constructs were injected in single injections as follows: pCS2+/LifeAct-GFP, 6ng/µl; pCS2+/H2B-RFP, 20 ng/µl; pCS2+/angulin-1-3xGFP, 10 ng/µl; pCS2+/SF9-3xGFP, 8 ng/µl; pCS2+/CA-MYPT, 100 ng/µl. All p-αtubulin and p-nectin probes were injected at 7.5 ng/µl. For LSR knockdown experiments, 40 ng of LSR MO was injected (Gene Tools). For LSR MO and CA-MYPT experiments, the LSR MO and CA-MYPT were co-injected with the fluorescent nuclear marker H2B-RFP at the 16 and 32-cell stages to specifically target intercalating MCCs.

**Immunostaining**

Immunostaining for LSR was performed as previously described[30]. MO injected embryos were fixed in 2% PFA in 1xPBS for 2h and then washed in 1xPBS for 20 minutes before blocking overnight at 4°C in a blocking solution, 1xTBS with CasBlock™ (ThermoFisher). The rabbit anti-LSR antibody was used in a 1:50 dilution and incubated for two days at 4°C. Embryos were then washed in blocking solution overnight and incubated with an anti-rabbit Alexa 647 secondary antibody (ThermoFisher) using a 1:250 dilution. After incubating for 24h, embryos were then washed in the blocking solution overnight and then added phalloidin-448 (Abberior) for 2h at room temperature. Embryos were imaged immediately after finishing the protocol to guarantee staining quality.

**Confocal Microscopy**

All live imaging of early neurula-stage (Nieuwkoop and Faber, NF stage 13) *Xenopus* embryos was performed with confocal laser scanning inverted microscopes Zeiss LSM880 and LSM980 with Airyscan2 detector equipped with a 40x C-Apochromat W autocorr M27 (NA = 1.2, working distance = 0.28 mm) or with a 25x Objective LD LCI Plan-Apochromat autocorr M27 (NA = 1.2, working distance = 0.28 mm). Embryos had the vitelline membrane removed and were left to recover for 30 minutes before being mounted in a drop of 1% Ultra Low Melting Point Agarose (Sigma) prepared in 1/3X MMR. Embryos were then live imaged at room temperature. Movies of intercalating MCCs were acquired with a 40x objective with 30, 60 or 120 second intervals in Figs. 1,2,4 and 5. Large field of view movies of developing embryonic epithelia (Fig. 3) were acquired with 25x objective at 300 second intervals.

## Image Processing and Analysis

### Filopodia analysis pipeline and Filopodia/Cortex intensity analysis

The pipeline for the epithelial vertices and filopodia analysis was performed as follows. First, the 3D stacks of the MCC intercalation process were resliced (x-z direction), and maximum projection was applied to all the slices at every time point. Then the vertices on either side of the cell contour were tracked using the Manual tracking plugin (https://imagej.nih.gov/ij/plugins/track/track.html) in Fiji. Using the position coordinates, the distance between left and right vertices were calculated. To mark the cell outline, the base of the filopodia was manually traced and the Region of Interests (ROIs) were recorded (Supplementary Fig. 1c). The cell tip given by the highest point of cell outline and the vertex positions were used to calculate the distance between the cell and the vertices. Similarly, the cell tip and the center of line connecting left and right vertices were used to compute the distance between the cell and epithelial surface. With the cell outline as reference, the contour was shifted above and below with appropriate thickness to match the filopodia (ROI #3) and cell cortex (ROI #2) respectively (Supplementary Fig. 1c). Background intensity values were collected for correction (ROI#1) (Supplementary Fig. 1c). In order to identify the prominent filopodia, we normalized the mean intensity of every pixel along the filopodia contour with the mean intensity of cortex, and highlighted those pixels with the values above "1" in magenta (Fig. 1f,g). Further, to assess the bias between the presence of filopodia closer to the left and right vertices, we introduced a parameter called filopodia index. The filopodia index is defined as the sum of intensities along the filopodia contour divided by the length of contour. With the midpoint as reference, we split the contour into left and right, which allows us to calculate the left and right filopodia index (Supplementary Fig. 1 d,e). All the steps in this pipeline were performed using scripts written in Fiji (for data collection) and in MATLAB R2017b (for analysis and plotting). The first part of the pipeline was also adapted to determine the accumulation of F-actin and myosin at the leading edge of intercalating cells for Fig. 5b, Supplementary Figs. 4 and 7b,d. Different ROIs were extracted for quantitative analysis: 1 - background, 2 - trailing edge of intercalating cell, 3 - cortex and 4 - filopodia of intercalating cell (see Supplementary Fig. 4d). F-actin and myosin intensities at the leading edge (cortex or filopodia) were normalized to the trailing-edge of the cell. For Supplementary Fig. 4f, individual intensity measurements for each timepoint were pooled for statistical comparison of filopodia/cortex mean intensity between control and LSR depleted MCCs.

### Intercalation success quantification

Intercalation success was qualitatively defined by a cells' ability to expand its apical domain using the control cells as the reference. Control and depleted cells were pooled from the same embryos. Cell blockage, cell death and cell disappearance are manually quantified. Cells are quantified as blocked if they fail to expand their apical domains, as dying if they undergo apoptosis and fragmentation and as disappearing if they move out of the superficial layer without undergoing obvious apoptosis and fragmentation (Supplementary Fig. 3a-d).

**LSR/F-actin intensity quantification**

LSR accumulation at the epithelial vertices in LSR-immunostained embryos was manually quantified in Fiji. 3.5 μm diameter circular ROIs were drawn over the position of the vertices determined using the F-actin/phalloidin channel. These ROIs were then used to collect the mean intensity values for F-actin and LSR. Intensity distributions for control and LSR depleted cells were plotted using OriginPro 2020. In Supplementary Fig. 2g, the LSR accumulation at the filopodia was manually quantified by drawing a ROI along the filopodium and using the Plot Profile tool in Fiji. Grey intensity values were then extracted for both channels and normalized to a cytoplasmic region. Normalized intensity values were plotted from base to tip.

**Image segmentation and cell tracking**

Sum intensity projection images of the SF9-3xGFP were segmented using the Cellpose segmentation algorithm[45]. We used the pretrained model cyto2 with the following conditions: cell diameter 50, flow threshold 0 and cell probability threshold 6.0. Mistakes in the segmentation masks were then corrected using the Tissue Analyzer drawing function. Tissue Analyzer was used to track cells and junctions and to obtain morphological information on both[46]. Cells at the edges that failed to be properly segmented were excluded from subsequent analysis.

**Vertex number, Myosin II and Sum of line tensions quantification**

Raw myosin intensity values at the junctions (using Maximum intensity projection images) were extracted using Tissue Analyzer[46]. Myosin II intensity values were normalized to the total cell intensity and then used for the computation of the sum of line tensions.

**Higher fold-Intercalation, remodelling percentage and junction length quantification**

Tracking of segmented movies from Tissue Analyzer were used to quantify the type of intercalation. Each intercalation was manually tracked to determine the local organization of the site of the expanding MCC and scored depending on the conformation of the neighboring cells (3-, 4-, 5-, 6- or 7-fold/neighbours). Higher-fold intercalations were then manually scored for concurrence with remodelling events. remodelling events were scored by tracing back the original conformation of the vertices before MCC apical expansion. Only remodelling events simultaneous to MCC intercalation were scored as remodelling. From these events the junction length distribution involved in forming higher-order vertices was calculated with t=0 min marking the onset of MCC intercalation. The other represented timepoints (t=-20 min, t=+20 min, t=+40 min,

t=+60 min) are aligned to t=0 min. Only events where one MCC can unequivocally be tracked throughout the remodelling event were quantified.

**Junction and MCC Myosin II quantification**

Junction length was manually tracked by tracing the junction outline using the segmented line function from ImageJ (exemplified in Supplementary Fig. 6a). Intensity values were extracted from ROIs with 1 µm thickness. Junction myosin II and actin mean intensities were normalized and corrected for photobleaching by dividing the shrinking junction intensity by the mean intensity of a non-dynamic junction (a junction that does not significantly change length over time). Similar analysis was performed to obtain the myosin and actin intensity at the leading edge of the MCC in Supplementary Fig. 5d.

**Vertex pulling quantification**

3D hyperstacks of MCC interacting with epithelial vertices were resliced (x-z direction) and projected to obtain a maximum intensity projection. The signal corresponding to LSR-3xGFP or LSR-GFP was filtered using a median filter with a 2 pixel kernel and then manually thresholded to obtain the epithelial vertex outline. Using the Analyze Particles tool of Fiji we obtain ROIs outlining the vertex. The Feret's diameter (corresponding to the longest possible distance between two points in the ROI) was then used as a measure of vertex length (referred to as Feret length in the figures to avoid possible confusion). MCC F-actin intensity was measured inside the ROI as a proxy for the contact between MCC and the epithelial vertice and it was normalized to the trailing edge of the cell. For average MCC F-actin intensity and Feret length, pulling events are aligned to the Feret length maximum (T=0), which we understand to be the peak of pulling at the vertex (Supplementary Fig. 2c).

**Statistics**

Statistical analysis was performed using the Origin2020 software. Non-parametric Mann-Whitney U-tests were used for analysis of statistical significance in Supplementary Figs. 3f and 4f. One-way Analysis of Variances (ANOVA) with Tukey's test was used to compare difference junction lengths with the reference junction length (junction length at the onset of MCC intercalation, T=0) in Supplementary Fig. 5e. The experiments were not randomized, and no statistical method was used to select sample size. Result reproducibility was confirmed by performing independent experiments and all experiments have a minimum of three replicates except for Supplementary Fig. 3f.

**Conflicts of interests**

The authors declare that they have no conflicts of interest.

# FIGURE 1

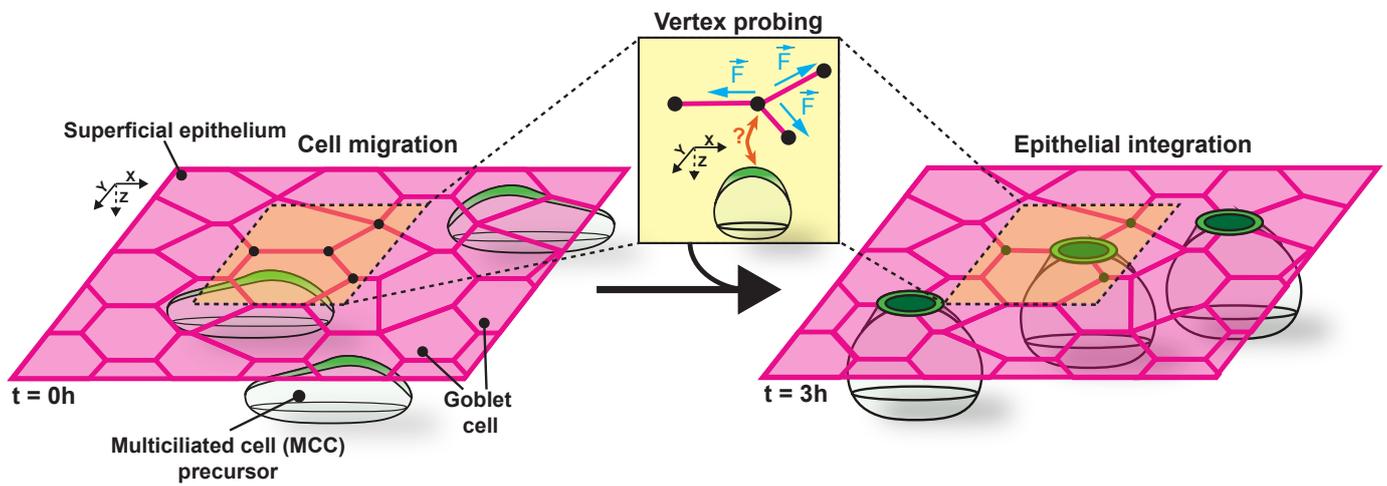

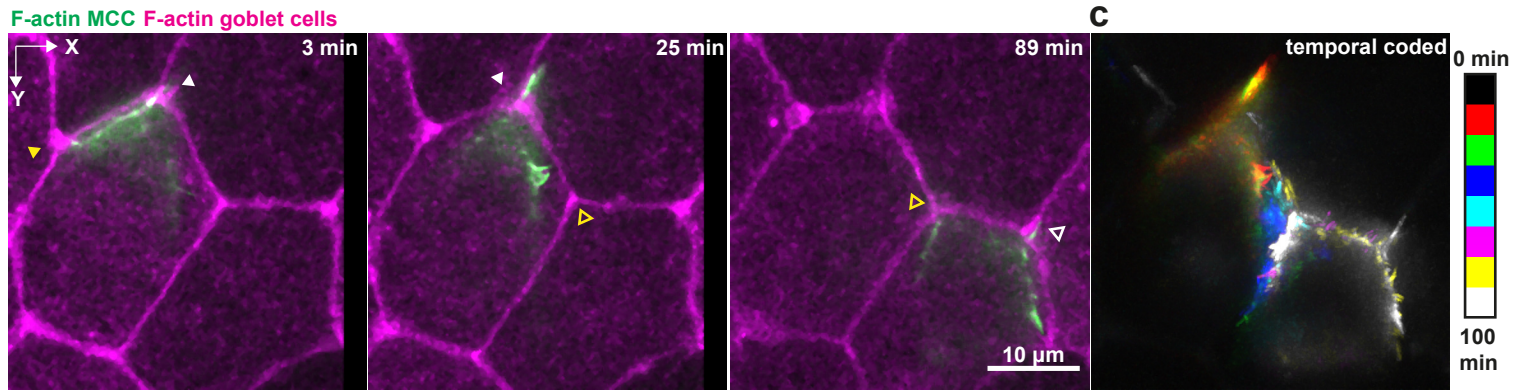

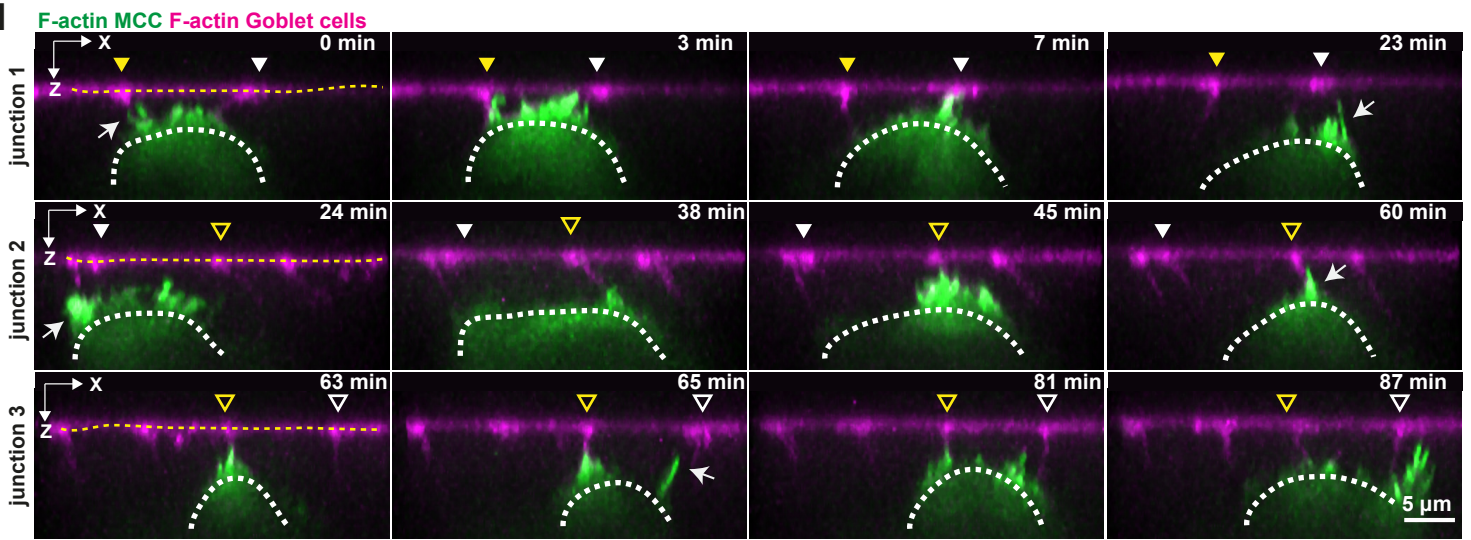

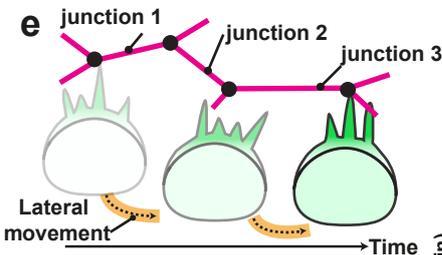

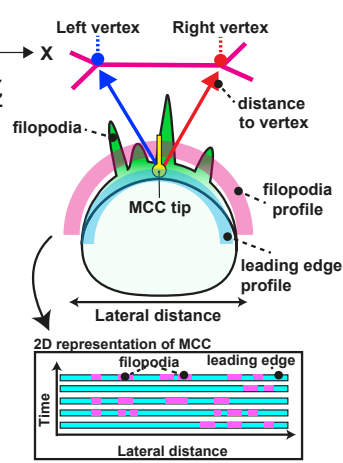

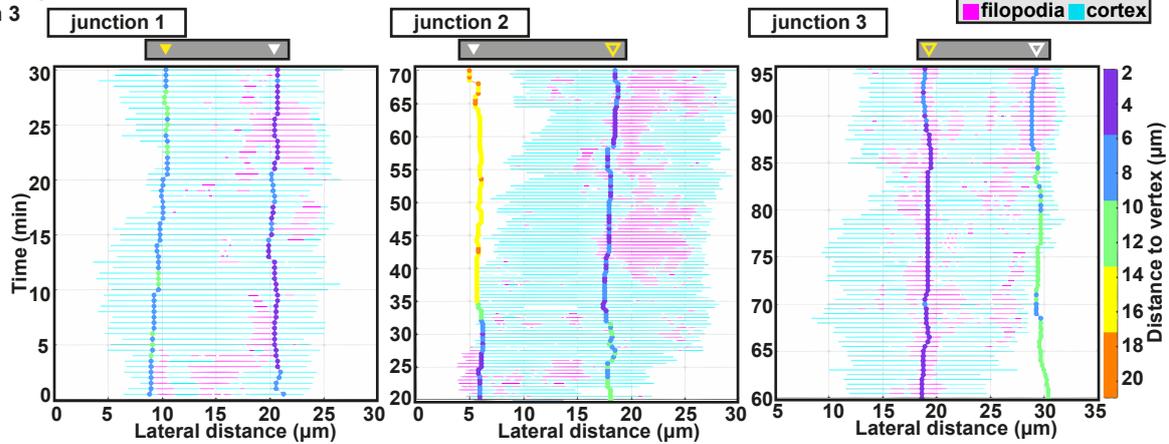

**Figure 1: Actin-based protrusions guide MCC intercalation. a,** Schematics representing multiciliated cell (MCC) intercalation. MCCs migrate into the superficial epithelium (t=0) and then integrate in the tissue at specific positions, the epithelial vertices formed by the goblet cells (t=3h). Inset depicting vertex probing: epithelial vertices (black dots) form hotspots of mechanical tension as connecting junctions (magenta) pull on the vertex (blue arrows). **b-g,** Dynamics of MCC probing (MCC expressing α-tubulin:LifeAct-GFP (green), goblet cells expressing nectin:utrophin-RFP (magenta)). Yellow and white arrowheads, with and without fill, mark the position of different vertices and white arrows mark F-actin protrusions. **b,** Image sequence of MCC moving in between the overlying goblet cells. Scale bar: 10 μm. **c,** Temporal-colored XY projection of cell in b. **d,** Orthogonal (XZ) projections used for filopodia dynamics analysis. White dotted lines outline the MCC contour and yellow dotted lines outline the top of the superficial epithelium. Scale bar: 5 μm. **e,** Schematics representing lateral movement of intercalating MCCs. **f,** Schematics representing F-actin protrusion analysis pipeline (see Methods). **g,** Relative position of F-actin protrusions extended by a single MCC and the overlaying epithelial vertices (vertical tracks, color-coded for distance) during lateral movement. Intercalating MCC (cyan) moves along different junctions as it extends protrusions (magenta).

# FIGURE 2

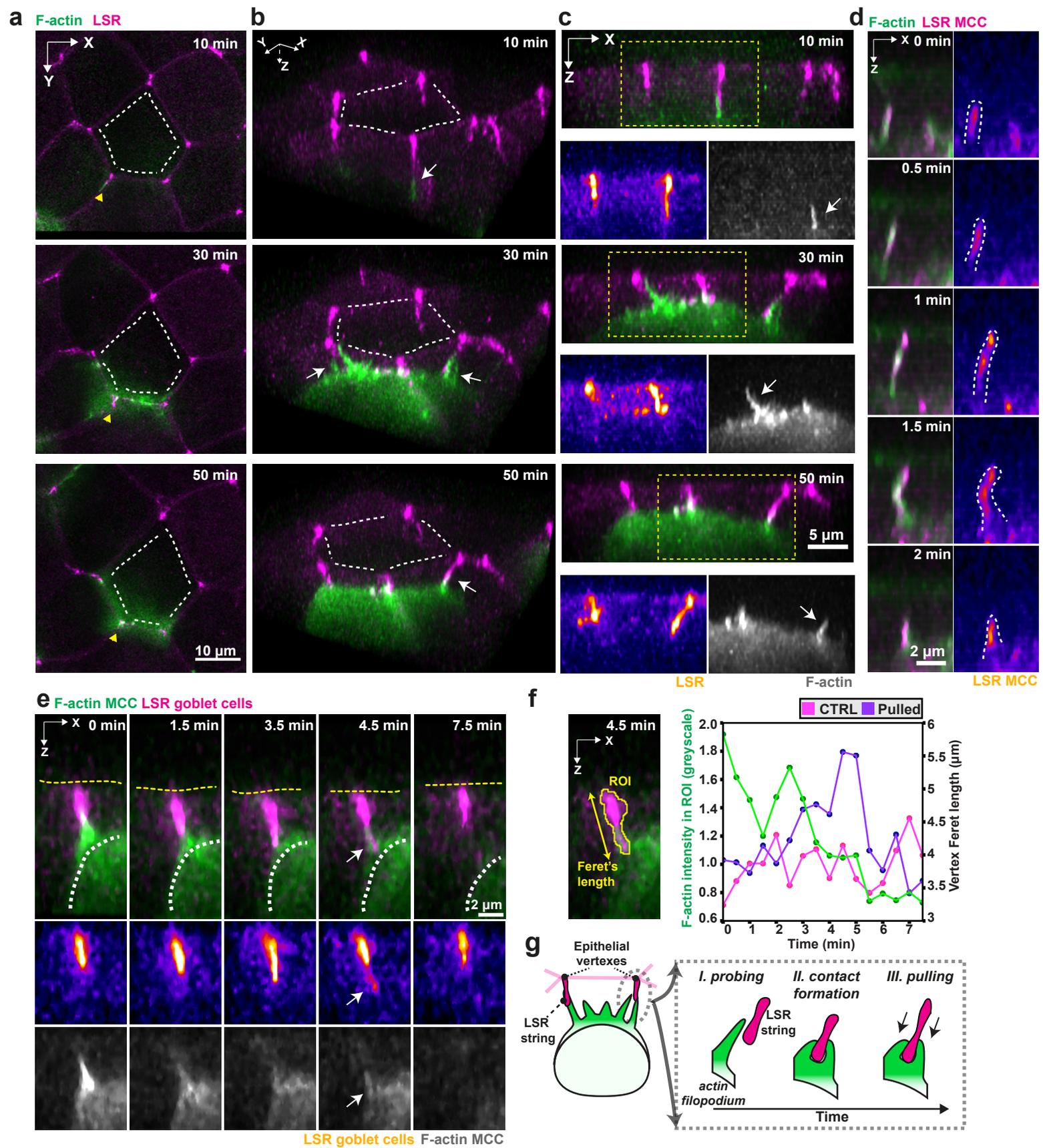

**Figure 2: LSR controls interaction between MCCs and insertion sites. a-e,** Filopodia closely interact with epithelial vertices as cells emerge into the superficial epithelium. Epithelial vertices are labeled with LSR-x3GFP or LSR-GFP (magenta in composite and fire as a separate channel) and MCCs by α-tubulin:LifeAct-RFP (green in composite and grey as a separate channel). **a,** Image sequence of intercalating MCC interacting with overlying vertices. Scale bar: 10 μm. White lines outline overlaying junctions. Yellow arrowheads depict orientation used for 3D rendering in b. **b,** 3D rendering of intercalating MCC forming contacts with different vertices (marked by white arrows). White lines outline overlaying junctions. **c,** Orthogonal (XZ) projections depicting close attachment between filopodia (marked by white arrows) and vertices. Yellow box marks inset for separate channels. Scale bar: 5 μm. **d,** Close-up of LSR (magenta) recruitment to F-actin based protrusions (green). Scale bar: 2 μm. **e**, Orthogonal (XZ) projections depicting F-actin protrusion pulling on the epithelial vertex (marked by white arrows). White dotted line outlines the MCC contour and yellow dotted line outlines the apical surface of the superficial epithelium. Scale bar: 2 μm. **f**, Quantification of vertex pulling. MCC F-actin intensity (green) and vertex length (purple) during one vertex pulling and retraction event (pulled) and for a non-pulled vertex (control, in magenta). **g,** Schematics representing MCC probing and vertex pulling.

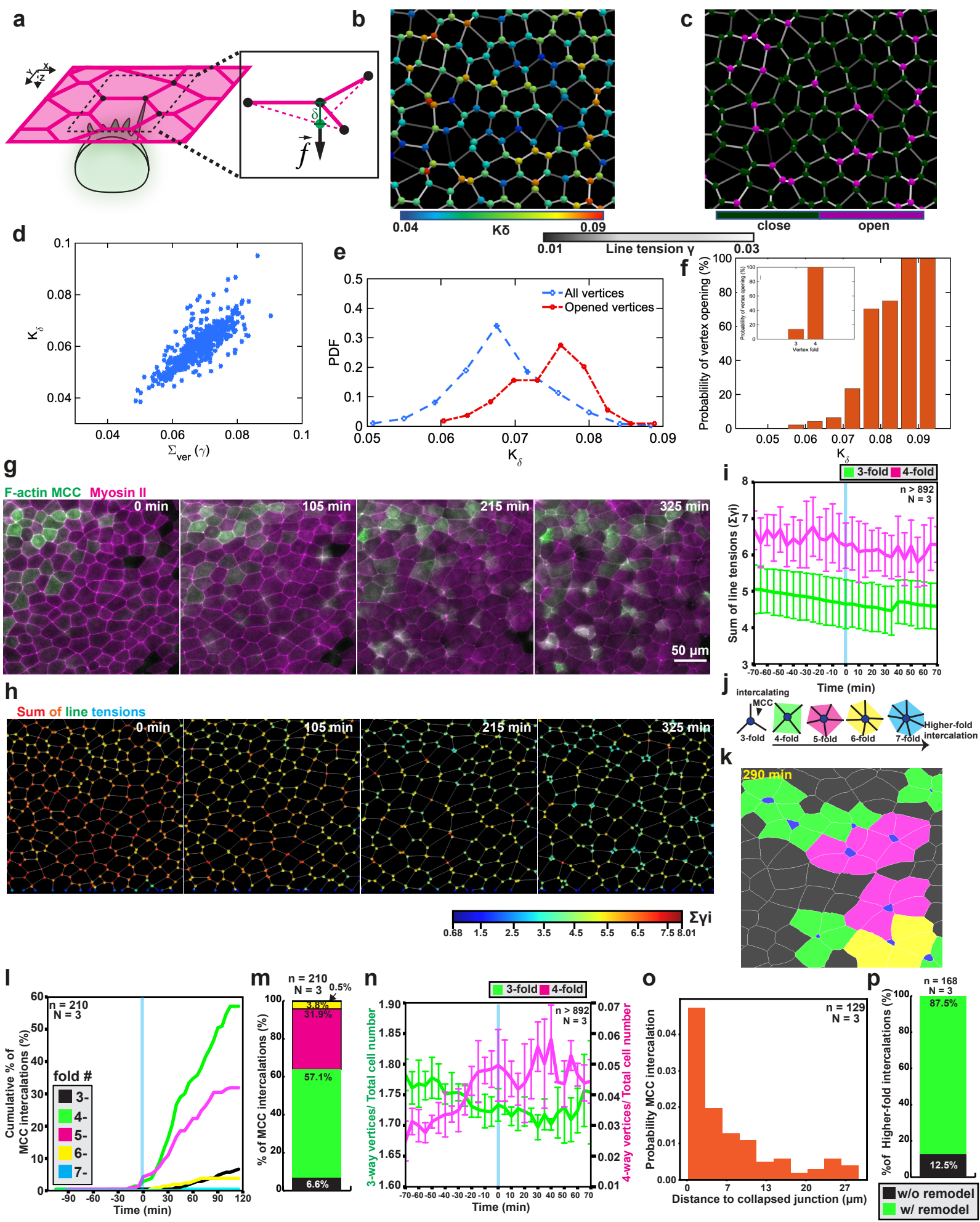

**Figure 3: MCCs predominantly intercalate at higher-fold vertices. a**, Schematics representing the out-of-plane force (*f*) exerted by intercalating MCCs on the epithelial vertex, which induces an out-of-plane displacement (δ). **b**, Representative snapshot of the simulated cellular network. The colormap on edges indicates line tension, while the vertices are colored according to their stiffness parameter. **c**, Representative snapshot of the cellular network illustrating propensity of vertices to open upon intercalation. The cell insertion events are probed at each vertex, one-at-a-time, and successful (failed) insertions are marked by purple (dark green) vertices. **d**, The vertex stiffness increases with increasing sum of line tensions at each vertex. **e**, Probability density function of the vertex stiffness distributions for all vertices in the simulated cellular network (blue) and for only the vertices that insertion events were successful (red). **f**, Probability of vertex opening for varying vertex stiffness values. The inset shows the probability of opening for 3- versus 4-fold vertices. **g,** Snapshots of developing mucociliary epithelium throughout intercalation. Intercalating MCCs and Myosin-II in goblet cells are labeled with α-tubulin:LifeAct-GFP (green) and the myosin nanobody SF9-3xGFP (magenta), respectively. Scale bar: 50 μm. **h**, Sum of line tensions across time. Sum of line tensions are color-coded from low (blue) to high tensions (red). **i**, Sum of line tensions plots for 3-fold vertices (green) and 4-fold vertices (magenta) during intercalation. Error bars represent SD. **j-m,** Quantification of MCC intercalations according to neighbour number. **j,** Scheme representing higher-fold intercalations, which are colored according to neighbor number. **k,** Segmented image used for analysis. Higher-fold intercalations are colored according to scheme j. Apical domains of intercalating MCCs are marked in blue. **l,** Cumulative percentage of MCC intercalations across time. T=0 marks the onset of MCC intercalation (defined as 1% addition of new cells). (n=210 cells from N=3 embryos). **m,** Total percentage of MCC intercalations. (n=210 cells from N=3 embryos). **n,** Evolution of 3-fold vertices (green) and 4-fold vertices (magenta) number across time, (from n>892 cells, N=3 experiments). T=0 (blue line) marks the onset of MCC intercalation (defined as 1% addition of new cells). Error bars represent SEM. **o**, Probability density of MCC intercalation as a function of distance to location of the closest edge collapse. **p**, Relative percentages of intercalation concomitant with (green) and without junction remodeling (black) (n=168 cells from N=3 embryos).

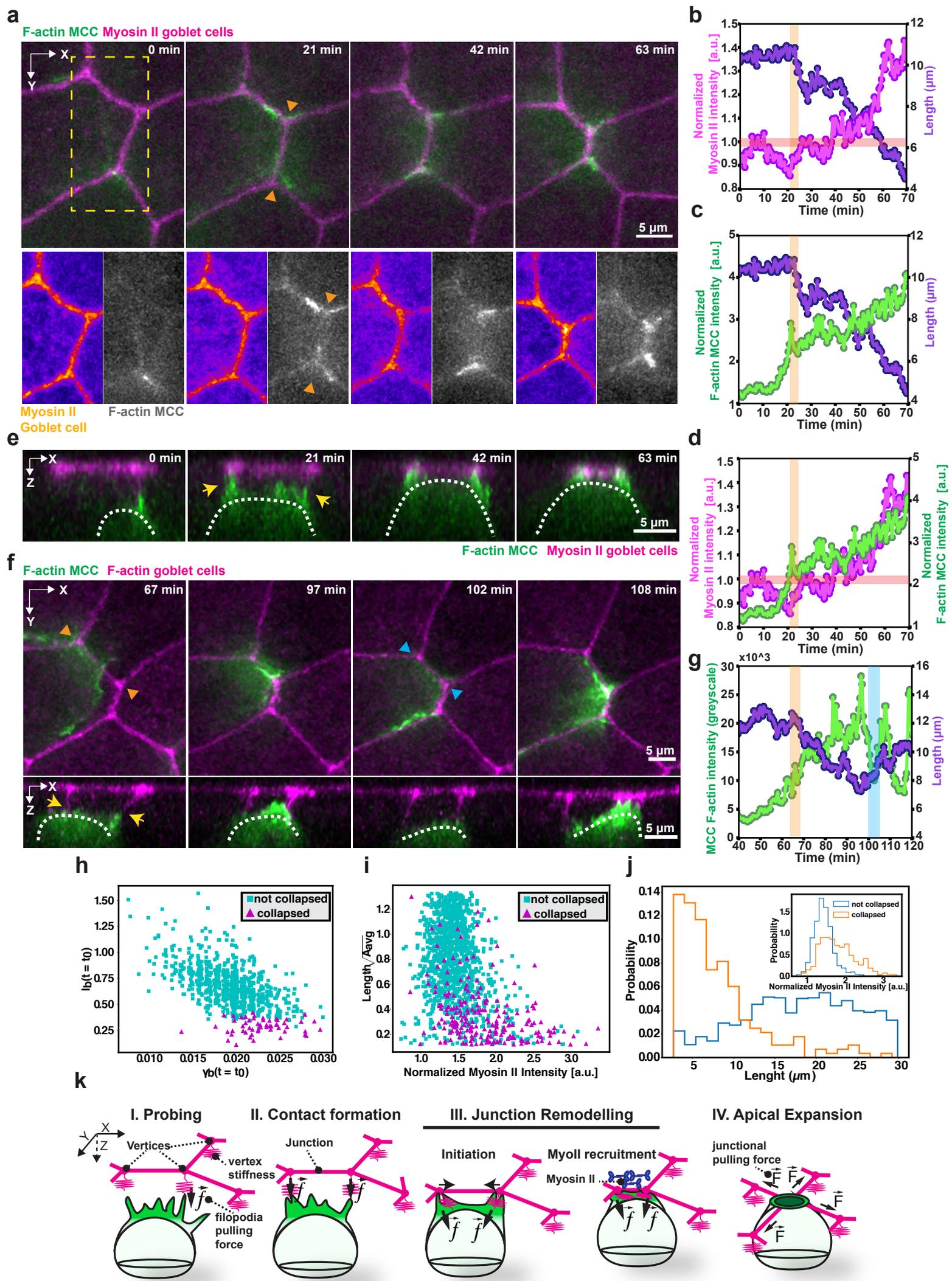

**Figure 4: MCCs remodel the neighboring epithelium to enable intercalation.** Intercalating MCCs and Myosin-II in goblet cells are labeled with α-tubulin:LF-GFP (green) and the myosin nanobody SF9-3xGFP (magenta), respectively. Orange arrowheads mark onset of junction collapse and cyan arrowheads mark loss of contact with the vertices. Orange bars mark onset of junction remodelling and cyan bars mark loss of contacts. **a-e,** Junction remodeling concurrent to MCC intercalation. **a,** High temporal resolution imaging of remodelling during intercalation. Scale bar: 5 μm. Yellow box marks inset for separate channels. **b,** Normalized junctional myosin-II intensity (magenta) and length (purple) during collapse. **c,** Normalized MCC F-actin intensity (green) and length (purple) during collapse. **d,** Normalized junctional myosin-II (magenta) and normalized MCC F-actin (green) intensities during collapse. **e,** Orthogonal (XZ) projections of MCC during junction remodelling. Yellow arrows depict contact with the epithelial vertices. Scale bar: 5 μm **f,** Retraction as part of the remodelling process. Scale bar: 5 μm. White dotted line outlines the MCC contour. Yellow arrows depict contact with the epithelial vertices. **g,** MCC F-actin intensity (green) and length (purple) during collapse and retraction. **h,i**, Stability-diagram of the edge collapse in the junction length-tension phase space from **h** simulation and **i** experiments. **j** Distribution of junction length and tension for collapsed and non-collapsed junctions in the experiments showing a stronger sensitivity of junction collapse to the initial length compared to the tension (shown in the inset). **k,** Schematics representing the multi-step, cooperative process of junction remodelling.

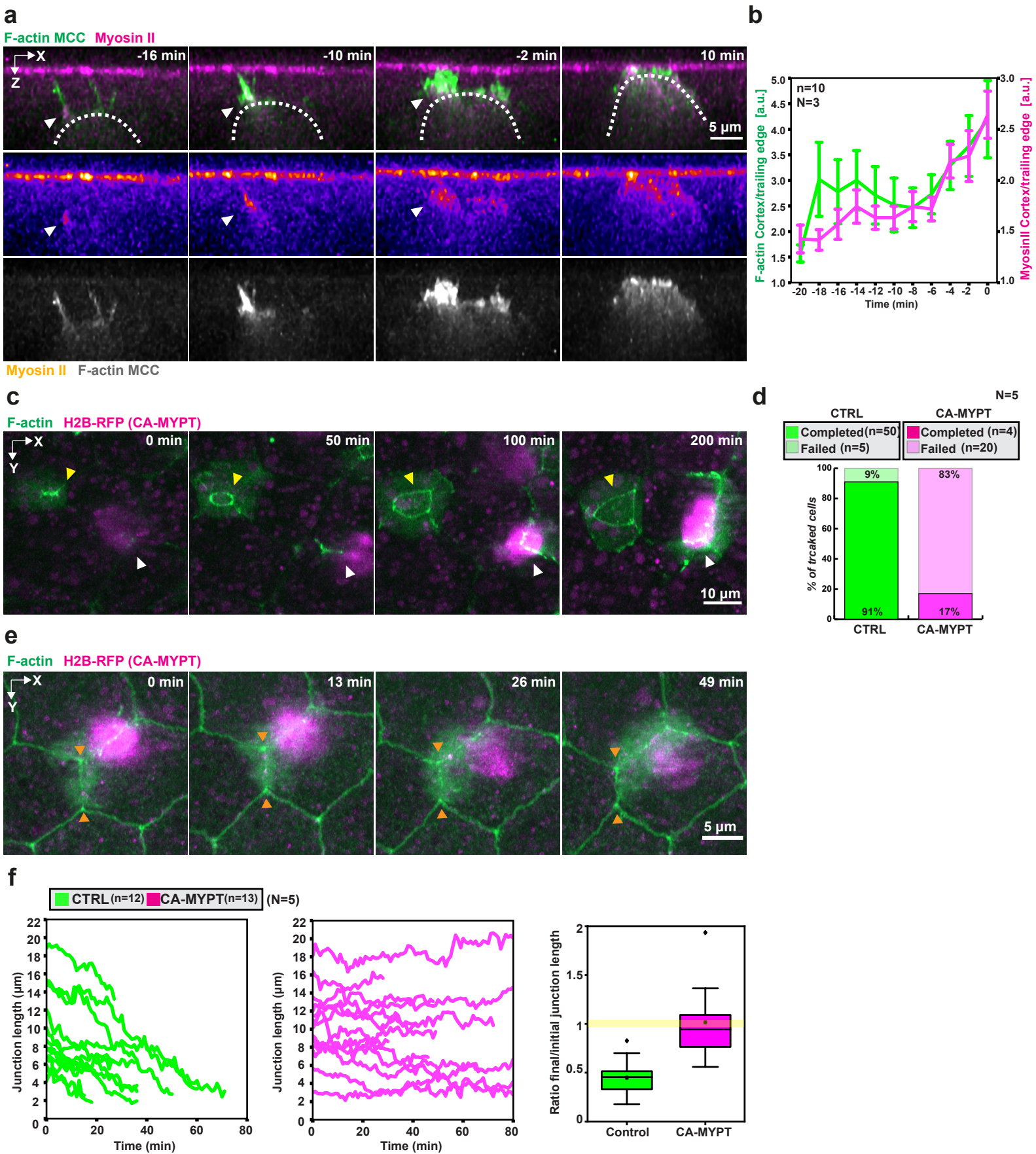

**Figure 5: Myosin II is required for cell intercalation and junction remodelling. a,** Orthogonal (XZ) projections of myosin recruitment in MCC during intercalation. Intercalating MCCs and Myosin-II are labeled with α-tubulin:LifeAct-GFP (green in composite, grey in separate channel) and the myosin nanobody SF9-3xGFP (magenta in composite, fire in separate channel), respectively. White arrowheads depict myosin recruitment. Scale bar: 5 µm. **b,** Normalized myosin-II intensity (magenta) and normalized F-actin intensity (green) at the leading edge of intercalating MCC. T=0 marks the last tracked frame during intercalation (n=10 cells from N=3 embryos). Error bars represent SEM. **c,** Image sequence of control MCC (yellow arrowhead) and CAMYPT-overexpressing MCCs (white arrowhead) during cell intercalation. Scale bar: 10 µm. **d,** Quantification of intercalation success rates (nWT= 55 cells, nCAMYPT = 24 cells, N=5 experiments). **e,** Image sequence of CAMYPT-overexpressing MCC attempting junction remodelling (orange arrowheads). Scale bar: 5 µm. **f,** Junction length tracking for control and CAMYPT-overexpressing MCCs (nWT= 12 junctions, nCAMYPT = 13 junctions, N=5 experiments). Boxplots of initial/final junction length in control and CA-MYPT MCCs. Yellow line indicates no overall remodelling (initial/final length=1).